\documentclass[fleqn,usenatbib]{mnras}

\usepackage{newtxtext,newtxmath,longtable}
\usepackage{xspace,upgreek,caption,subcaption}

\usepackage[T1]{fontenc}

\DeclareRobustCommand{\VAN}[3]{#2}
\let\VANthebibliography\thebibliography
\def\thebibliography{\DeclareRobustCommand{\VAN}[3]{##3}\VANthebibliography}

\usepackage{graphicx}	
\usepackage{amsmath}	

\newcommand{\frb}{FRB~20220912A\xspace}
\newcommand{\rone}{FRB~20121102A\xspace}
\newcommand{\rthree}{FRB~20180916B\xspace}
\newcommand{\gcfrb}{FRB~20200120E\xspace}
\newcommand{\rsixtyseven}{FRB~20201124A\xspace}
\newcommand{\ronetwin}{FRB~20190520B\xspace}

\title[Milliarcsecond Localisation of \frb ]{Milliarcsecond Localisation of the Hyperactive Repeating \frb }

\author[D. M. Hewitt et al.]{
Danté M. Hewitt$^{1}$,\thanks{E-mail: d.m.hewitt@uva.nl}
Shivani Bhandari$^{2,3,1,4}$,
Benito Marcote$^{3}$,
Jason W.~T. Hessels$^{2,1}$,
Kenzie Nimmo$^{5}$,
 \newauthor
Franz Kirsten$^{6}$,
Uwe Bach$^{7}$,
Vladislavs Bezrukovs$^{8}$,
Mohit Bhardwaj$^{9}$,
Richard Blaauw$^{2}$,
Justin D. Bray$^{10}$,
\newauthor
Salvatore Buttaccio$^{11}$,
Alessandro Corongiu$^{12}$,
Marcin P. Gawro\'nski$^{13}$,
Marcello Giroletti$^{11}$,
Aard Keimpema$^{3}$,
\newauthor
Giuseppe M. Maccaferri$^{11}$,
Zsolt Paragi$^{3}$,
Matteo Trudu$^{12}$,
Mark P. Snelders$^{2,1}$,
Tiziana Venturi$^{11}$,
Na Wang$^{14}$,
\newauthor
David R.~A. Williams-Baldwin$^{10}$,
Nicholas H. Wrigley$^{10}$,
Jun Yang$^{6}$ and
Jianping P. Yuan$^{14}$
\\
$^{1}$Anton Pannekoek Institute for Astronomy, University of Amsterdam, Science Park 904, 1098 XH, Amsterdam, The Netherlands\\
$^{2}$ASTRON, Netherlands Institute for Radio Astronomy, Oude Hoogeveensedĳk 4, 7991 PD Dwingeloo, The Netherlands\\
$^{3}$Joint institute for VLBI ERIC, Oude Hoogeveensedĳk 4, 7991 PD Dwingeloo, The Netherlands\\
$^{4}$CSIRO Space and Astronomy, Australia Telescope National Facility, PO Box 76, Epping, NSW 1710, Australia\\
$^{5}$MIT Kavli Institute for Astrophysics and Space Research, Massachusetts Institute of Technology, 77 Massachusetts Ave, Cambridge, MA 02139, USA\\
$^{6}$Department of Space, Earth and Environment, Chalmers University of Technology, Onsala Space Observatory, SE-439 92, Onsala, Sweden\\
$^{7}$Max-Planck-Institut für Radioastronomie, Auf dem Hügel 69, 53121 Bonn, Germany\\
$^{8}$Engineering Research Institute Ventspils International Radio Astronomy Centre of Ventspils University of Applied Sciences, Inženieru street 101,\\Ventspils, LV-3601, Latvia\\
$^{9}$McWilliams Center for Cosmology, Department of Physics, Carnegie Mellon University, Pittsburgh, PA 15213, USA\\
$^{10}$Jodrell Bank Centre for Astrophysics, Department of Physics and Astronomy, The University of Manchester, Manchester M13 9PL, UK\\
$^{11}$INAF-Istituto di Radioastronomia, Via Gobetti 101, 40129, Bologna, Italy\\
$^{12}$INAF-Osservatorio Astronomico di Cagliari, via della Scienza 5, I-09047, Selargius (CA), Italy\\
$^{13}$Institute of Astronomy, Faculty of Physics, Astronomy and Informatics, Nicolaus Copernicus University, Grudziadzka 5, 87-100 Toru\'n, Poland\\
$^{14}$Xinjiang Astronomical Observatory, CAS, 150 Science 1-Street, Urumqi, Xinjiang 830011, China
}

\date{Accepted XXX. Received YYY; in original form ZZZ}

\pubyear{2015}

\begin{document}
\label{firstpage}
\pagerange{\pageref{firstpage}--\pageref{lastpage}}
\maketitle

\begin{abstract}
We present very-long-baseline interferometry (VLBI) observations of the hyperactive repeating \frb using the European VLBI Network (EVN) with an EVN-Lite setup. We detected 150 bursts from \frb over two observing epochs in October 2022. Combining the data of these bursts allows us to localise \frb to a precision of a few milliarcseconds, corresponding to a transverse scale of less than 10\,pc at the distance of the source. The precision of this localisation shows that \frb lies closer to the centre of its host galaxy than previously found, although still significantly offset from the host galaxy's nucleus. On arcsecond scales, \frb is coincident with a persistent continuum radio source known from archival observations, however, we find no compact persistent emission on milliarcsecond scales. The persistent radio emission is thus likely to be from star-formation in the host galaxy. This is in contrast to some other active FRBs, such as \rone and \ronetwin.  
\end{abstract}




\section{Introduction}

Fast radio bursts (FRBs) are flashes of coherent radio emission that have durations of microseconds to seconds \citep[for a recent review see][]{petroff_2022_aarv}. Some of them are known to repeat \citep{spitler_2016_natur}. While more than 2\,000 unique sources of FRBs have been detected to date \citep{chime_2023_apj}, less than 50 have been localised to a host galaxy\footnote{The FRB Community Newlsetter (Volume 04, Issue 12) reported 44 host galaxies \url{https://hosting.astro.cornell.edu/research/frb/news/}}. Precise localisations of FRBs are key to understanding their origins and for using them as astrophysical and cosmological probes.
While arcsecond precision is normally sufficient to identify a host galaxy robustly \citep{eftekhari_2017_apj}, sub-arcsecond localisations are key to identifying the exact galactic and stellar neighbourhoods in which FRB sources reside \citep[e.g.,][]{tendulkar_2021_apjl}.

Magnetars are widely favoured as the engines powering FRBs, given the high burst rate and millisecond timescales associated with FRBs \citep[e.g.,][]{ravi_2019_natas,li_2021_natur,nimmo_2021_natas}, as well as the detection of exceptionally bright radio bursts from the Galactic magnetar SGR~1935+2154 that were coincident with an X-ray burst \citep{bochenek_2020_natur,chime_2020_natur_galacticfrb}. Nonetheless, the diverse properties and environments of FRB sources suggests that a single magnetar progenitor model may be overly simplistic \citep[e.g.,][]{kirsten_2022_natur}.

FRB sources show a wide range of activity rates, from a few hyper-active repeaters to apparent one-off events that constitute 97\% of the currently known sources \citep{chime_2023_apj}. The lack of obvious bi-modality in the burst rates suggests that one-off sources may be capable of repeating \citep{chime_2023_apj}, but statistically significant differences in burst properties between repeaters and non-repeaters suggest they may be distinct \citep{pleunis_2021_apj}. Another possibility is that a single source model is capable of producing multiple burst types \citep{hewitt_2023_mnras,snelders_2023_natas}. Both repeaters and (apparent) non-repeaters have been localised to a wide variety of host galaxies, with no clear distinction in galaxy type \citep{gordon_2023_apj,bhardwaj_2023_arxiv}. There is an overall trend towards star-forming galaxies, with notable exceptions \citep{sharma_2023_apj}.  

The global galactic properties of an FRB host are only indirectly informative about the source's nature. More directly, we can study the local environment of FRB sources via time-variable propagation effects \citep[e.g.,][]{michilli_2018_natur} and precision localisation coupled to high-resolution imaging \citep[e.g.,][]{mannings_2021_apj}. Ideally, radio localisations should have $<100$\,mas uncertainty, in order to maximize the degree to which one can zoom-in on their local environment. Thus far, only five repeaters, and as yet no non-repeaters, have been localised to milliarcsecond precision \citep{marcote_2017_apjl,marcote_2020_natur,kirsten_2022_natur,nimmo_2022_apjl,bhandari_2023_apjl}.

The first detected repeater, \rone, was localised to a low-metallicity star-forming dwarf galaxy \citep{chatterjee_2017_natur,tendulkar_2017_apjl}. These observations also showed that \rone was spatially consistent with a faint persistent radio source (PRS). Follow-up observations by the EVN \citep[European VLBI Network; ][]{marcote_2017_apjl} enabled a milliarcsecond localisation of the bursts and concrete association between the bursts and PRS (a projected linear separation of $\lesssim$40\,pc). 
They also showed that the PRS is compact on sub-pc scales, and hence cannot be due to local star formation. Rather, it may be a hyper-nebula powered by the FRB source, or a low-luminosity active galactic nucleus \citep{marcote_2017_apjl}. The precision of this \rone localisation further enabled characterisation of the local environment using the \textit{Hubble Space Telescope} \citep[\textit{HST}; ][]{bassa_2017_apjl}, which revealed that \rone\ is inside, but slightly off center ($\sim200$\,pc) from a knot of star-formation in its host galaxy. Together with the source's extreme and highly variable Faraday rotation measure (RM), this supports the case for a young magnetar progenitor \citep[e.g.,][]{metzger_2019_mnras}. 

Thereafter, the repeating, and periodically active \citep{chime_2020_natur_periodicactivity}, \rthree\ was localised by the EVN to a nearby massive spiral galaxy \citep{marcote_2020_natur}.
The precision of the EVN localisation allowed for the association of the FRB source with the apex of a relatively large, apparently \mbox{v-shaped} star-formation region, but also ruled out the presence of a PRS, distinguishing it from the other known and localised repeater at the time. Follow-up \textit{HST} observations showed that \rthree\ is located slightly offset ($\sim250$\,pc) from the nearest knot of star formation --- suggesting that it is a neutron star  formed by a runaway massive star, or perhaps an older neutron star in a binary system \citep{tendulkar_2021_apjl}.

The hyper-active repeater \rsixtyseven \citep[e.g.,][]{lanman_2022_apj}, was first localised to arcsecond precision by the Australian Square Kilometre Array Pathfinder \citep[ASKAP; ][]{day_2021_atel_14515}, Very Large Array/realfast \citep[VLA; ][]{law_2021_atel}, and upgraded Giant Metrewave Radio Telescope \citep[uGMRT; ][]{wharton_2021_atel_14538}. The host galaxy was found to be star-forming, dusty and an order-of-magnitude more massive than the hosts of other repeaters at the time, bridging the gap between the hosts of repeaters and apparent non-repeaters \citep{ravi_2022_mnras}. The VLA (in D-configuration) and uGMRT detected unresolved, persistent radio emission at radio frequencies of 3 and 9\,GHz \citep{ricci_2021_atel}, and 300\,MHz \citep{wharton_2021_atel_14529}, respectively. Follow-up observations with the VLA (in C-configuration) at 22\,GHz, however, resolved this emission, disqualifying it as a compact PRS and showing that the radio emission is more likely due to star formation \citep{piro_2021_aa}. Milliarcseond localisation with the EVN \citep{nimmo_2022_apjl} found no evidence for compact radio emission coincident with the burst position, supporting the notion that the previously detected emission is of extended nature and that \rsixtyseven is embedded in a region of star formation. The milliarcsecond localisation also enabled deeper, high-resolution radio and optical studies with the VLA and \textit{HST}, respectively, leading to the conclusion that the FRB source formed \textit{in situ} \citep{dong_2023_arxiv}.

Using the raw voltage data of three bursts, the Canadian Hydrogen Intensity Mapping Experiment FRB project \citep[CHIME/FRB; ][]{chime_2018_apj} localised \gcfrb\ to the outskirts of the M81 spiral galaxy complex (at a luminosity distance of 3.6\,Mpc) with a 90 per cent confidence interval of $\simeq14$\,arcmin$^2$ \citep{bhardwaj_2021_apjl}. Follow-up observations by \cite{kirsten_2022_natur} confirmed that \gcfrb is indeed associated with the M81 galactic system and, surprisingly, coincident with a globular cluster. This finding challenged theories that advocate that all FRBs originate from young magnetised neutron stars formed via core collapse SNe (supernovae). If \gcfrb is indeed such a magnetised neutron star, alternative formation channels need to be invoked: e.g., formation via binary merger or accretion-induced collapse of a white dwarf \citep{kremer_2021_apjl}.

Finally, \ronetwin, discovered by the Five-hundred-meter Aperture Spherical Telescope (FAST), was localised to a dwarf host galaxy at a $z=0.241$ using the VLA \citep{niu_2022_natur}. VLA observations identified a potential PRS with a flux density of $\sim 200 \upmu$Jy at 3\,GHz. Recent observations with the EVN have confirmed the compact PRS nature by constraining the transverse size of the source to be $<9$\,pc. These observations have also showed that the FRB source and the PRS are consistent with being co-located within $\leq80$\,pc --- consistent with the hypothesis that a single central engine must power both the bursts and the PRS \citep{bhandari_2023_apjl}.

The primary focus of this paper, a hyper-active repeater called \frb, was discovered by CHIME/FRB \citep{mckinven_2022_atel}. CHIME/FRB reported a position of RA (J2000): 347.29(4)$^{\circ}$, Dec (J2000): +48.70(3)$^{\circ}$ (90 per cent uncertainty errors), and a dispersion measure (DM) of 219.46(4) pc\,cm$^{-3}$ for this source. The FRB position lies somewhat outside of the Galactic plane: $l = 106.1^{\circ}$, $b = -10.8^{\circ}$. The expected scattering timescale from the Galactic interstellar medium (ISM) along this line-of-sight is a moderate $2.6\,\upmu$s (at 1\,GHz) according to the NE2001 Galactic electron density model \citep{cordes_2002_arxiv}. The Deep Synoptic Array (DSA-110) localised \frb to a host galaxy, PSO~J347.2702+48.70, at $z = 0.0771(1)$ \citep{ravi_2023_apjl}. The host galaxy has a stellar mass of approximately $10^{10}\,$M$_{_{\odot}}$ and a star-formation rate of $\gtrsim$\,0.1\,M${_{\odot}}$\,yr$^{-1}$, making it unremarkable compared to some other known host galaxies of repeaters \citep{gordon_2023_apj}. \frb is the most active FRB known to date, with FAST detecting as many as 390 bursts per hour \citep{zhang_2023_apj}. Interestingly, the local environment of the source also appears to be clean, as the RM of the bursts has been stable around zero for a period on the order of months \citep[e.g.,][]{feng_2023_arxiv,zhang_2023_apj,hewitt_2023_mnras}. 

Here we present the interferometric localisation of 150 bursts detected from \frb\ with EVN-Lite observations in October 2022. Section~\ref{sec:observations} outlines the technical details of our observations. We describe our search pipeline, localisation procedures and the measurement of burst properties in Section~\ref{sec:analysis}. Finally, our main conclusions are presented and placed in the context of other FRBs in Section~\ref{sec:discon}.

\section{Observations}
\label{sec:observations}
\begin{figure}
    \centering
    \includegraphics[width=0.48\textwidth]{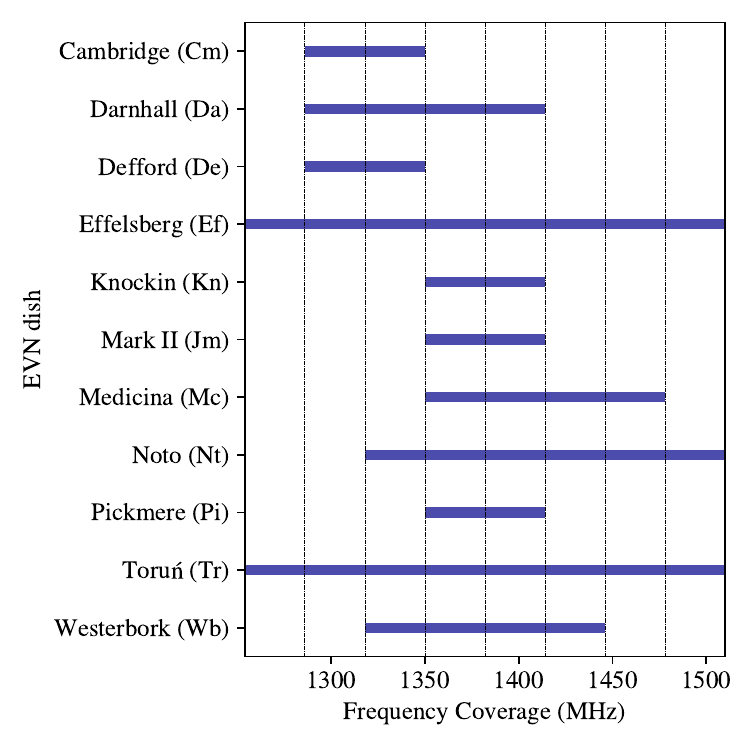}    
    \caption{Our PRECISE observations spanned a total bandwidth of 256\,MHz from 1254--1510\,MHz. The frequency coverage of each of the EVN dishes in the array is shown in the plot by the horizontal bars, while the dashed vertical lines indicate the edges of the sub-bands. Noto observes from 1350--1606\,MHz, but only the range below 1510\,MHz, where there is overlap with other stations, is correlated.}     
    \label{fig:freq_coverage}
\end{figure}

We observed \frb in three observing runs in October 2022 as part of the ongoing FRB VLBI localisation programme called PRECISE (Pinpointing REpeating ChIme Sources with EVN dishes; PI: Kirsten). The first observation (Epoch~1; EVN project code EK051G; PRECISE code PR249A) was conducted on 22 October 2022 from 00:00--04:46~UT, and utilised an \textit{ad hoc} array of 11 EVN and eMERLIN dishes: Cambridge, Darnhall, Defford, Effelsberg, Knockin, Jodrell Bank Mark II, Medicina, Noto, Pickmere, Toru\'{n}, and Westerbork. The second observation (Epoch~2; EK051H; PR247A), was conducted from 24 October 2022 21:00~UT to 25 October 2022 02:00~UT. Westerbork did not participate in the second observation and the array consisted of the remaining 10 dishes from Epoch~1. The third observing run (Epoch~3; PR248A), was conducted from 26 October 2022 23:00 UT to 27 October 2022 04:30 UT. During this run we used the 11 aforementioned dishes as well as the Onsala 25-m telescope. In the first two observations, we pointed the array to a sky position of RA=23$^{\rm{h}}$09$^{\rm{m}}$05.49$^{\rm{s}}$ Dec=+48$^{\circ}$42$^{\prime}$25.6$^{\prime\prime}$, which is the position of the initial localisation determined using the DSA-110 \citep[][]{ravi_2022_atel_15693}. We note that this position is $5.8^{\prime\prime}$ offset from the final reported DSA-110 position, RA=23$^{\rm{h}}$09$^{\rm{m}}$04.9$^{\rm{s}}$ Dec=+48$^{\circ}$42$^{\prime}$25.4$^{\prime\prime}$ \citep{ravi_2023_apjl}, but still well within the primary beam of all EVN dishes. In Epoch~3 we pointed to the updated DSA-110 position. Observations for Epoch~1 and 2 were carried out at a central frequency of 1.4\,GHz with bandwidth ranging from 64--256\,MHz for the different antennas, and we recorded dual-polarization raw voltage data in a circular basis with 2-bit sampling at all the participating stations in VDIF \citep{whitney_2010_ivs} format. The frequency coverage was not identical between individual dishes and is illustrated in Figure~\ref{fig:freq_coverage}. For Epoch~3 we observed with a similar set-up but at higher frequencies (4798--5054\,MHz). We provide a more concise overview of Epoch~3 as no bursts were detected during this higher-frequency observation \citep{kirsten_2022_atel}.

For Epoch~1, our observations interleaved target scans of 5.75\,min on \frb\ and scans of 1.5\,min on a nearby ($3.0^{\circ}$ offset) phase calibrator source, J2311+4543, resulting in phase referencing cycles with a duration of approximately 7.25\,min. Every fifth iteration we also observed another nearby source, J2314+4518 ($0.6^{\circ}$ offset from the phase calibrator), for 3.5\,min to be used as an interferometric check source. This check source is used to estimate the absolute astrometric uncertainty and potential amplitude losses that might have been introduced during phase referencing. A 5\,min scan was scheduled on J1327+4326 to use as a fringe finder and bandpass calibrator.  Finally, the pulsar B2111+46 was also observed for 5\,min to verify the integrity of our data for the burst search and single pulse analyses with Effelsberg. A similar strategy was followed in Epoch~2 and 3, but in Epoch~2 the phase calibrator J2311+4543 was also used as a fringe finder and bandpass calibrator, the check source was replaced with J2327+4754 ($3.4^{\circ}$ offset from the phase calibrator) and B0329+54 was used as a test pulsar. In Epoch~3 we used J2308+4629 as a phase calibrator, J2327+4911 as a check source again ($4.1^{\circ}$ offset from J2327+4911), J2311+4543 as the fringe finder and B2020+28 and B0540+23 as test pulsars.

\section{Analysis and Results}
\label{sec:analysis}
\subsection{Search for Bursts}

\begin{figure*}
    \centering
    \includegraphics[width=1\textwidth]{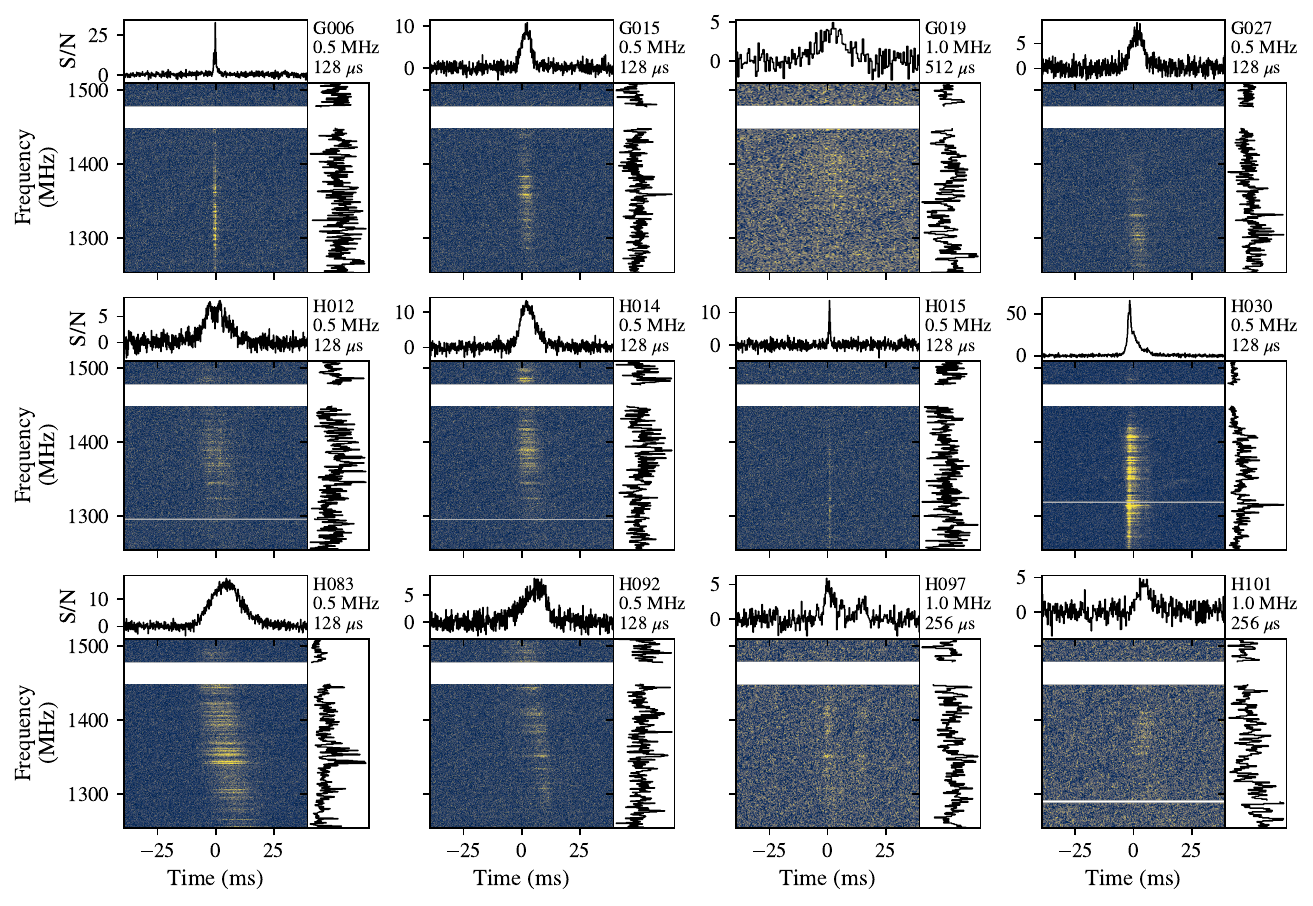}    
    \caption{This sub-sample of bursts detected from \frb with the Effelsberg dish illustrates the diversity in burst duration and the complex morphology seen in some high-S/N bursts. All bursts have been dedispersed using a DM of 219.37\,pc\,cm$^{-3}$. In each thumbnail, the main panel shows the dynamic spectrum of the burst. The top panel shows the frequency-averaged time profile (averaged over the spectral extent of the burst), while the side panel shows the time-averaged frequency spectrum.  For visual purposes the bursts have been averaged in time and frequency, and the plotted time and frequency resolutions are shown in the top-right corner of each thumbnail. Horizontal white lines in the dynamic spectrum indicate channels that have been masked due to the presence of RFI.}     
    \label{fig:family_plot}
\end{figure*}

We searched the raw voltage data from Effelsberg for bursts, using the pipeline previously described in detail in \cite{kirsten_2021_natas}. In short, the raw voltage data were converted to Stokes~I filterbanks with time and frequency resolutions of 64$\,\upmu$s and 62.5\,kHz, respectively, using \texttt{digifil} \citep{vanstraten_2011_pasa}. We then used the GPU-accelerated transient detection software \texttt{Heimdall}\footnote{\url{https://sourceforge.net/projects/heimdall-astro/}} to search a DM range of $169-269$\,pc\,cm$^{-3}$ for FRB candidates that are above a signal-to-noise ratio (S/N) of 7. The resulting candidates were classified by the machine learning convolutional neural network \texttt{FETCH} \citep{agarwal_2020_mnras}, using the models A and H\footnote{Empirical tests that we ran showed that models A and H performed best in terms of completeness and number of false positives.}. All candidates for which these models assigned a $>0.5$ probability of the burst being astrophysical in origin were manually inspected. We detected a total of 45 and 105 bursts in the first and second observation, respectively. No bursts were detected in the 5-GHz data from Epoch~3. A sub-sample of the bursts is shown in Figure~\ref{fig:family_plot}.

\subsection{Correlation of Interferometric Data}

The PRECISE data were correlated in numerous passes at the Joint Institute for VLBI ERIC (JIVE) in the Netherlands (EVN correlation proposal EK051; PI: Kirsten), making use of the software correlator \texttt{SFXC} \citep{keimpema_2015_exa}. For Epoch~1, the first pass was a delay-mapping correlation where three bursts and their bracketing phase calibrator scans were used to determine the burst position to an uncertainty of $\approx1$\,arcsec \citep[see detailed description in ][]{marcote_2020_natur}. The initial DSA-110 localisation \citep{ravi_2022_atel_15693} was used as the phase centre of the \frb\ target field. The correlation was done with $8\times32$-MHz sub-bands consisting of 64 spectral channels each and an integration time of 2\,s for the phase calibrator scans, while the target scans were manually gated according to the width of the bursts and coherently dedispersed to a DM of 219.46\,pc\,cm$^{-3}$. In the second pass, all the bursts were coherently dedispersed and correlated at a phase center determined by the delay-mapping. In order to maximise the S/N, gate widths were chosen around the arrival time of each burst. After the interferometric localisation of the \frb\ bursts described in the next sub-section, all target data were then re-correlated in a third and final correlation pass using this position as the phase center, in order to create a deep image to look for persistent radio continuum counterparts. We repeated the procedure for Epoch~2, but without the first delay-mapping pass (as the position was already known), and using the interferometric localisation from Epoch~1 as the phase center.

During these correlations we encountered a few technical issues that required resolution before finalising the analysis. The Earth Orientation Parameters
(EOPs) used in the correlation of the EK051H data were not properly updated in the correlator due to a failure of a software that pings NASA's Archive of Space Geodesy Data\footnote{\url{https://cddis.nasa.gov/Data_and_Derived_Products/Other_products/IERS_EOPs.html}}. This was discovered and corrected for the continuum data by re-correlating them with updated EOPs. However, the burst correlation still had outdated EOPs with a large discrepancy, notably in the UT1$-$UTC values. This initially introduced an offset of $\sim30$\,mas in RA for the position of the burst source between EK051G and EK051H. The offset was resolved internally at JIVE by performing an EOP correction on EK051H burst data. The correction applied a phase shift to the visibilities that corresponds to the delay difference that results from the different sets of parameters. These delays are approximated by using the IAU2000A precession and nutation model to calculate the celestial to terrestrial coordinate transformation matrix. The source code for this implementation is available online\footnote{\url{https://code.jive.eu/kettenis/correct_eops}}, and this feature will be added to \texttt{CASA} \citep{mcmullin_2007_aspc,vanbemmel_2022_pasp} in a future release.

\subsection{Burst Localisation}

The EVN data were calibrated using standard interferometric techniques in \texttt{AIPS} \citep{greisen_2003_assl} and \texttt{DIFMAP} \citep{shepherd_1994_baas}. Imaging was performed in \texttt{DIFMAP} and \texttt{CASA} v6.1.

After the correlated visibilities (in FITS-IDI format) were loaded into \texttt{AIPS}, we first applied the calibration table from the EVN \texttt{AIPS} pipeline that contains the parallactic angle correction and \textit{a-priori} gain correction, using the gain curves and system temperature measurements that the stations recorded during the observations. We also applied the \textit{a-priori} flagging table and bandpass calibration table. We then flagged the edges of sub-bands ($\approx15$\,per cent of the channels in total) and manually flagged data from the fringe finder scans that were contaminated by RFI (radio frequency interference). Ionospheric dispersive delays can have a significant impact on the calibration and localisation accuracy at milliarcsecond scales at L-band. To mitigate this we made use of the \texttt{VLBATECR} task in \texttt{AIPS} to correct for these delays. The task makes use of maps from the Jet Propulsion Laboratory of the total electron content (TEC) at the different EVN sites during the observations, and compensates for the dispersive delays accordingly. We used the fringe finder scans (J1327+4326 for Epoch~1 and J2311+4543 for Epoch~2), with Effelsberg as the reference antenna, to remove the phase jumps between sub-bands and phase slopes within sub-bands that are introduced because of the different signal paths for individual sub-bands. Next, a global fringe fit was performed to correct the phases of the entire observation for all calibrator sources as a function of both frequency and time. The solutions were manually inspected, and bad solutions were flagged.

Having applied the aforementioned calibration, we imaged the phase calibrator (J2311+4543) and check sources (J2314+4518 for Epoch~1 and J2327+4754 for Epoch~2) in \texttt{DIFMAP} using a cell size of 1\,mas and a natural weighting scheme (synthesized beam sizes of $\approx30\times40$\,mas). We were able to reproduce the positions of both check sources to a precision of $\lesssim$2\,mas compared to the expected positions from 5\,GHz maps of the sources in the RFC~2023B catalogue\footnote{\url{http://astrogeo.org/sol/rfc/rfc_2023b/rfc_2023b_cat.html}}.
For J2314+4518 we measure a positional offset of $\Delta\,\alpha\ = 0.1$\,mas and $\Delta\,\delta\ = 1.0$\,mas, and for J2327+4754, a positional offset of $\Delta\,\alpha\ = 1.6$\,mas and $\Delta\,\delta\ = 1.9$\,mas. The expected positions of J2314+4518 and J2327+4754 in the 5\,GHz maps have uncertainties of 1.06\,mas and 1.47\,mas, respectively. Taking these uncertainties into account, together with the difference in observing frequency, we conclude that our calibration was successful. We factor in these positional offsets when determining the FRB position and conservatively include the statistical uncertainty on the check source positions in the calculation of the final FRB positional uncertainty.

We also performed self-calibration to further improve our calibration solutions. We first imaged and self-calibrated the phase calibrator in \texttt{DIFMAP} to obtain the best possible model of the source. This model allowed us to improve the phases and amplitudes of the different antennas. The resulting model was imported into \texttt{AIPS} and was used to create a calibration table. Finally, we applied these calibration solutions to the target field of \frb\ and imaged the target (both the continuum data and burst data). We again used a cell size of 1\,mas and natural weighting.  

\begin{figure*}
    \centering
    \includegraphics[width=1\textwidth]{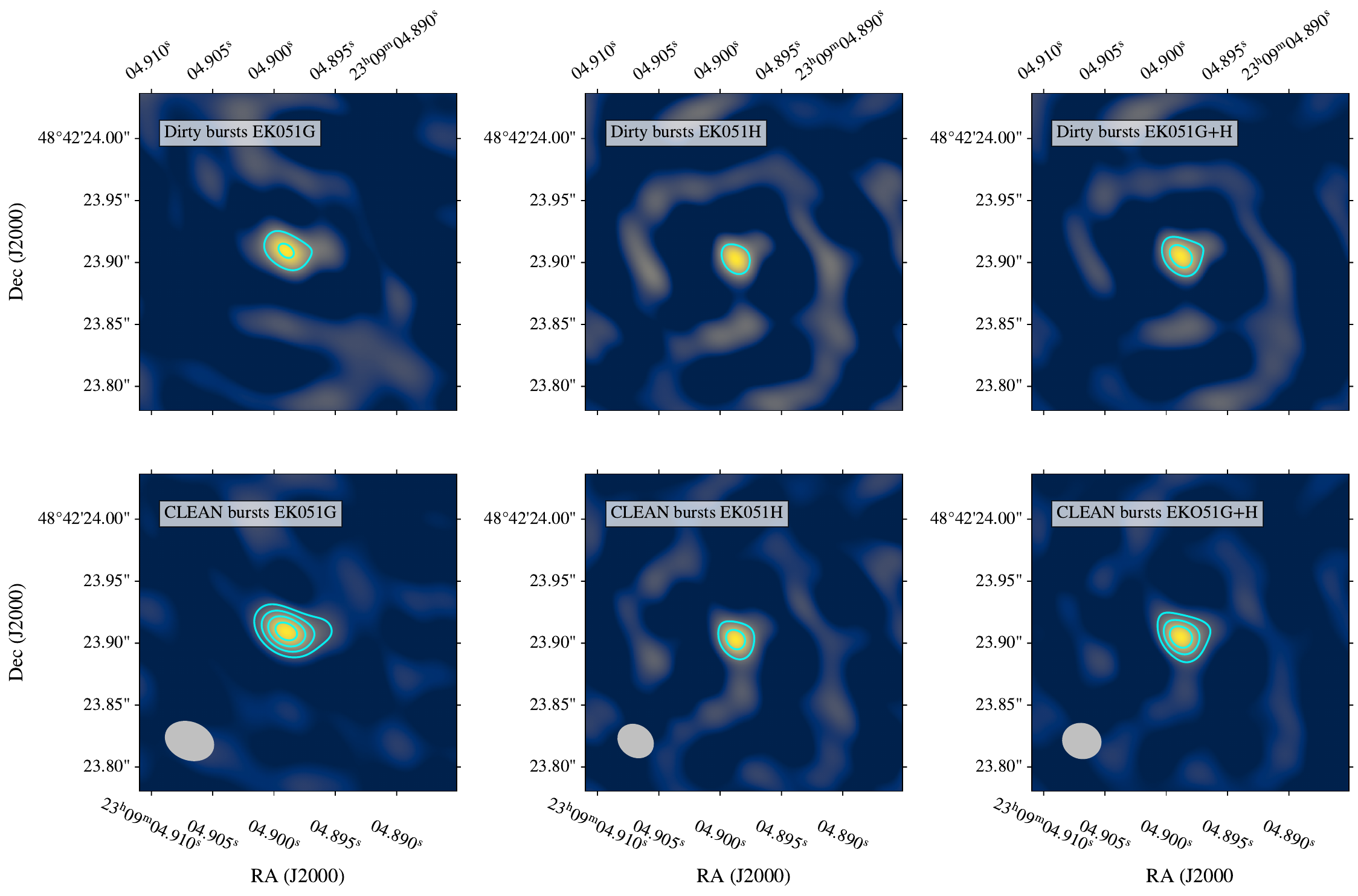}    
    \caption{In the top row, the dirty maps of the EVN 1.4\,GHz observation of the combined visibilities of the 45 bursts detected in Epoch~1 and 105 bursts detected Epoch~2 are shown in the left and middle panels, respectively. The combined visibilities of both epochs, i.e. all 150 bursts, are shown on the right. The cyan contours start at 4 times the RMS noise level of each image and increase by factors of 3.  In the bottom row the CLEAN images are shown with the synthesized beam displayed as a silver ellipse in the bottom left corner. }     
    \label{fig:burst_dirties}
\end{figure*}

We combined the visibilities of the 45 bursts detected in Epoch~1 and 105 bursts detected in Epoch~2 to create the dirty maps shown in Figure~\ref{fig:burst_dirties} with \texttt{CASA}. Taking into account the positional offset of our check sources, for Epoch~1 we find the position of the bursts from \frb to be RA (J2000) = $23^{\rm h}09^{\rm m}04.8990^{\rm s} \pm 3.4\,\rm{mas}$, Dec (J2000) = $48^\circ42^\prime23.9104^{\prime\prime} \pm 3.3~\mathrm{mas}$. For Epoch~2 we find a position of RA (J2000) = $23^{\rm h}09^{\rm m}04.8987^{\rm s} \pm 3.5\,\rm{mas}$, Dec (J2000) = $48^\circ42^\prime23.9053^{\prime\prime} \pm 3.5~\mathrm{mas}$. These positions are offset from one another by $\Delta\,\alpha\ = 3.5$\,mas and $\Delta\,\delta\ = 5.1$\,mas, but despite the total offset of 6.2\,mas, still consistent with one another to within $\sim$1$\sigma$ (the synthesized beam sizes are $29\times40$\,mas and $24\times30$\,mas for Epoch~1 and 2, respectively). The uncertainties we quote take into account multiple factors that are summed in quadrature: the statistical uncertainty derived from the shape and size of the synthesized beam normalized by the S/N ($\Delta$RA = 0.3\,mas, $\Delta$Dec = 0.3\,mas for Epoch~1 and $\Delta$RA = 0.4\,mas, $\Delta$Dec = 0.3\,mas for Epoch~2); the statistical uncertainty on the position of the phase calibrator, J2311+4543 (0.10\,mas); an estimate of the uncertainty from phase-referencing due to the angular separation between the phase calibrator and target \citep[$\sim$3\,mas; ][]{kirsten_2015_aa}; an estimate of the frequency-dependent shift in the phase calibrator position from the International Celestial Reference Frame (ICRF), here conservatively $\sim$1 mas \citep{plavin_2022_mnras}; and the statistical uncertainty on the positions of the interferometric check sources (1.06 and 1.47\,mas for Epochs~1 and 2, respectively). A more in-depth per epoch analysis is presented in Appendix~\ref{app:perepoch}. 

The final ICRF position of \frb from taking the average position of these two datasets (all 150 bursts) and, conservatively, the quadrature sum of the uncertainties, is: 
\\
\indent RA (J2000) = $23^{\rm h}09^{\rm m}04.8988^{\rm s}$; $\Delta$RA = 5\,\rm{mas}, \\
\indent Dec (J2000) = $48^\circ42^\prime23.9078^{\prime\prime}$; $\Delta$Dec =  5\,\rm{mas}. \\

\begin{figure*}
    \centering
    \includegraphics[width=1\textwidth]{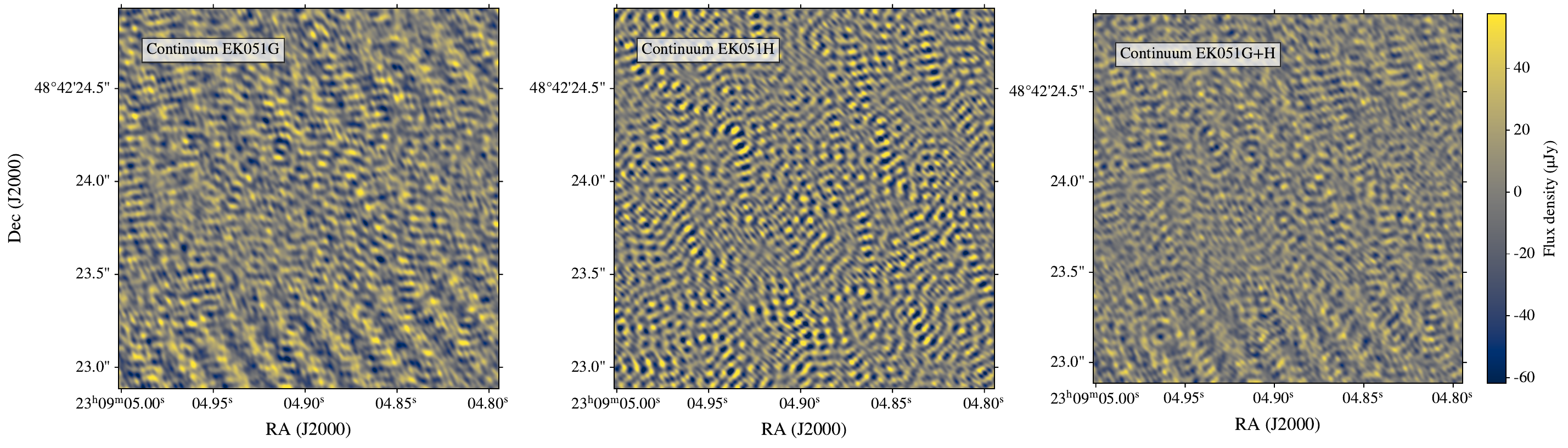}    
    \caption{The dirty maps (roughly speaking, the inverse Fourier transform of the visibilities) of the EVN 1.4-GHz observation of the $\sim2\times2\, \rm{arcsec}^2$ field surrounding \frb. We find no peak above 5$\sigma$ in these maps, thus placing a 5$\sigma$ upper limit of $L\approx$1.2$\times10^{28}$erg\,s$^{-1}$\,Hz$^{-1}$ on the brightness of a non-detectable PRS. The colour map is scaled to the dynamic range in the combined map.}     
    \label{fig:continuum}
\end{figure*}

We also imaged a $2\times2$\,arcsec$^2$ area surrounding the position of the bursts to search for a compact radio continuum counterpart (i.e., a PRS). Figure~\ref{fig:continuum} shows these dirty images. The images of the first and second epochs have an RMS of 21 and 24\,$\upmu$Jy\,beam$^{-1}$, respectively, while the combined data from both epochs have an RMS of 16\,$\upmu$Jy\,beam$^{-1}$. We find no evidence for any persistent radio emission on milliarcsecond scales surrounding \frb, ruling out the presence of a PRS above 5$\sigma$ significance ($L\approx$1.2$\times10^{28}$erg\,s$^{-1}$\,Hz$^{-1}$) in a region of $\sim2\times2\,$arcsec$^2$.

\subsection{Burst Properties}

For each burst, we used \texttt{digifil} to create filterbank files from the baseband data recorded by Effelsberg. The time and frequency resolution of these filerbanks were 64\,$\upmu$s and 62.5\,kHz, respectively. The bursts were then incoherently dedispersed to a DM of 219.37\,pc\,cm$^{-3}$. This DM value was determined by temporally aligning high-S/N, broadband microshots in the dynamic spectra of exceptionally bright bursts detected from \frb\ with the Nan\c{c}ay Radio Telescope \citep{hewitt_2023_mnras}. Optimising for structure by using \texttt{DM\_phase} \citep{seymour_2019_ascl}, with a bandpass filter on the fluctuation frequencies, yields similar results. Since we did not apply coherent dedispersion, we chose enough frequency channels to limit DM smearing in the lowest frequency channel to less than the time resolution of the data. After bandpass correction (subtracting the mean and dividing by the standard deviation of the off-burst noise on a per-channel basis) we applied a static mask at frequencies ranging from 1448--1477\,MHz, in addition to manually flagging channels that are contaminated by RFI. 

Using the function \texttt{curve\_fit} from the python package \texttt{scipy}, we fit a one-dimensional Gaussian function to the frequency-averaged light curve of each burst (only considering the spectral extent of the burst which was manually determined). We define the width of a burst as the FWHM of this Gaussian fit. To calculate the fluence of the burst, we first normalized the light curve and then integrated over the 3$\sigma$ extent of the 1D Gaussian fit, before multiplying with the radiometer equation \citep{cordes_2003_apj}. As is convention for Effelsberg observations, we assume a system temperature and gain of 20\,K and 1.54\,K\,Jy$^{-1}$, respectively. These values have an uncertainty of approximately 20\%, which propagates into the following energy calculations. We calculated the spectral energy density ($E_\nu$) as:
\begin{equation}
    E_\nu = \frac{4\pi F\Delta\nu D_{L}^2 }{\nu(1+z)},
\end{equation}
where $F$ is the fluence, $\Delta\nu$ is the spectral extent of the burst, $\nu$ is central observing frequency, and $D_L$ and $z$ are the luminosity distance (362.4\,Mpc) and redshift (0.0771) of the host galaxy of \frb, respectively \citep[][]{ravi_2023_apjl}.

\begin{figure*}
    \centering
    \includegraphics[width=1\textwidth]{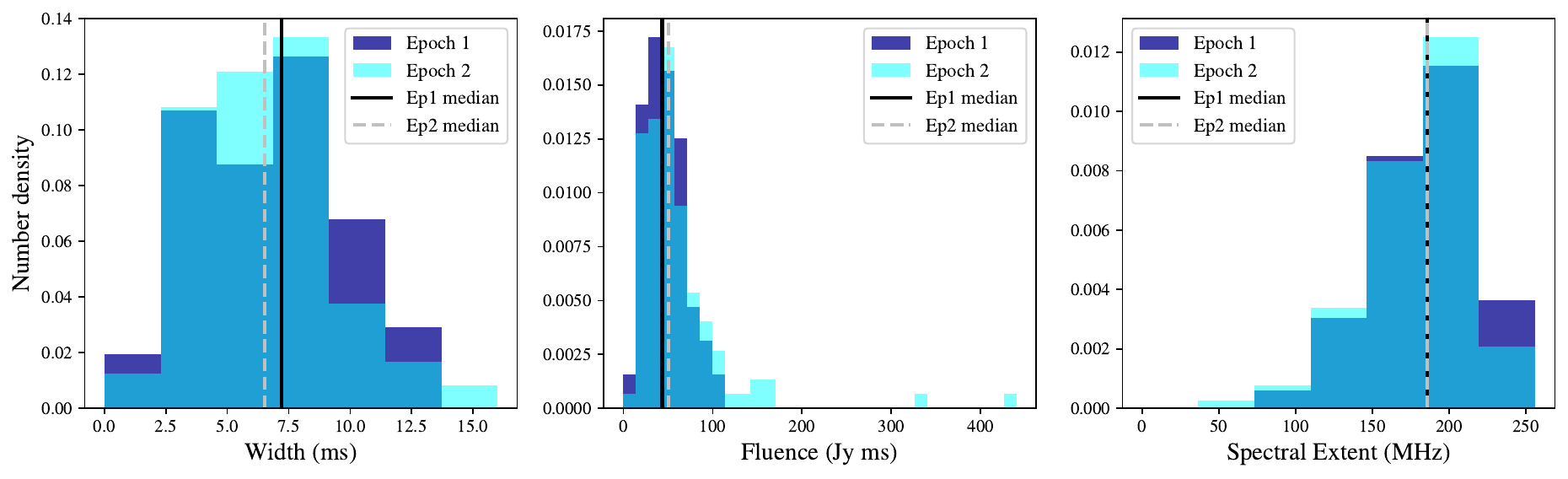}    
    \caption{The distribution of the temporal width, fluence, and spectral extent of the bursts we detected are shown in dark blue and cyan for Epoch 1 and 2, respectively. Note that the spectral extent distribution only reflects the observed spectral extent of bursts and are thus in many cases lower limits. The histograms have been normalised for each observation, so that the total area equals 1. Vertical lines indicate median values: solid black for Epoch~1 and dashed grey for Epoch~2. }     
    \label{fig:burst_properties}
\end{figure*}

These properties, as well as the times of arrival of the bursts, are tabulated in Table~\ref{tab:burst_properties}. Figure~\ref{fig:burst_properties} shows the normalized distribution of these properties, per epoch. The burst property distributions show little variation between the two epochs. Note that the spectral extent of the burst only considers the observed range, and is consequently often a lower limit as many bursts appear to have emission outside of our observing window. The median values for width and fluence of all bursts detected are 6.6\,ms and 47\,Jy\,ms, respectively.  

\section{Discussion and Conclusions} 

\label{sec:discon}

\begin{figure*}
    \centering
    \includegraphics[width=\textwidth]{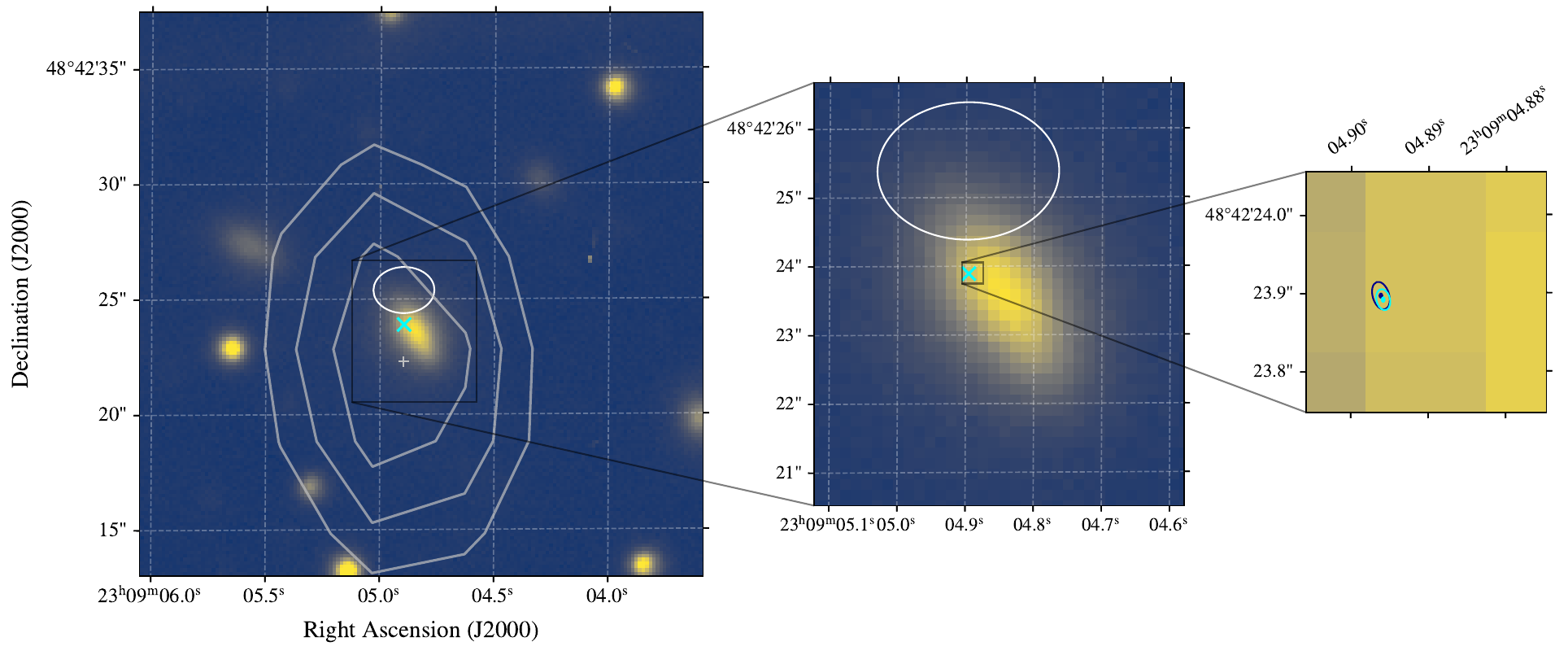}    
    \caption{Adapted from \citet{ravi_2023_apjl}, the background image is the deep optical image (R-band, obtained with Keck II/ESI; limiting magnitude R$\sim$26) of the host galaxy of \frb, PSO~J347.2702+48.7066. The grey plus-sign indicates the centroid of the catalogued radio continuum source, APTF~J230904+484222, while they grey contours of this source are at levels of 16, 20 and 24\,mJy. The approximate 90 per cent error ellipse of the DSA-110 localisation of \frb is over-plotted in white, while our EVN position is indicated by the cyan cross. Note that the uncertainty on our VLBI position is much smaller than the symbol size and the resolution of the optical image. The insets show consecutive zoom-ins on the EVN position of \frb. In the rightmost inset, the synthesized beams of the observations EK051G and EK051H are overplotted as dark blue and cyan ellipses, respectively, centered at the position found by combining the visibilities of all bursts at each epoch. The solid points at the center of the ellipses are roughly the size of 90 per cent error regions.}     
    \label{fig:optical}
\end{figure*}

In this paper we report the detection of 150 bursts from \frb using an \textit{ad hoc} EVN-Lite array of  dishes, which allowed us to localise this FRB source to a precision of a few milliarcsecond: RA (J2000) = $23^{\rm h}09^{\rm m}04.8988^{\rm s} \pm 5\,\rm{mas}$, Dec (J2000) = $48^\circ42^\prime23.9078^{\prime\prime} \pm 5~\mathrm{mas}$. \frb is now the sixth repeating FRB source to be localised to milliarcsecond precision using VLBI. We find that \frb\ is significantly closer to the optical centre of its host galaxy, PSO~J347.2702+48.70, compared to the earlier localisation presented by \citet{ravi_2023_apjl}, shown in Figure~\ref{fig:optical}. The transverse offset from the host galaxy center is $\approx0.8$\,kpc. The precision of this VLBI localisation corresponds to a physical length of less than 10\,pc at the redshift of the source, and this mas-level position will serve future high-resolution IR/optical/UV imaging with \textit{HST}, \textit{James Webb Space Telescope} \textit{JWST}, and the Extremely Large Telescope (ELT), which could reveal star-forming regions or other discrete sources coincident on parsec scales with the position of \frb.

\rone and \ronetwin are the only known repeating FRBs that exhibit a PRS \citep{marcote_2017_apjl,niu_2022_natur}, which may represent a hyper-nebula powered by the burst source \citep[e.g.,][]{sridhar_2022_apj}. These are also two of the four repeaters from which burst storms have been observed \citep[e.g.,][]{li_2021_natur,hewitt_2022_mnras,jahns_2023_mnras,niu_2022_natur}, the other two being \gcfrb \citep{nimmo_2023_mnras} and \rsixtyseven \citep[e.g.,][]{lanman_2022_apj,zhou_2022_raa}. As previously mentioned, in the case of \rsixtyseven there is persistent radio emission attributed to star formation that resolves out at higher angular resolution \citep[e.g.,][]{piro_2021_aa,nimmo_2022_apjl}, whereas \gcfrb resides in a globular cluster with no signs of persistent radio emission \citep{kirsten_2022_natur}.

The upper limit we have placed on the presence of a PRS for \frb is more than an order-of-magnitude below the luminosity of the PRSs associated with \rone\ and \ronetwin.
There exists a catalogued continuum radio source, APTF~J230904+484222, detected by the Westerbork Synthesis Radio Telescope Aperture Tile In Focus (WSRT-APERTIF) and located at RA (J2000) = $23^{\rm h}09^{\rm m}04.9^{\rm s} \pm 1.7\,\rm{arcsec}$, Dec (J2000) = $48^\circ42^\prime22.3^{\prime\prime} \pm 2.2\,\mathrm{arcsec}$, which is 1.6\,arcsec away from our VLBI position for \frb. APTF~J230904+484222 has a peak brightness of 0.27$\pm$0.04\,mJy\,beam$^{-1}$ at 1355\,MHz ($L\approx$3.9$\times10^{28}$erg\,s$^{-1}$\,Hz$^{-1}$). The contours and centroid of this source are overplotted on the optical image and VLBI position in Figure~\ref{fig:optical}. The centroid position does not coincide with the nucleus of the galaxy. We also explore 4$\times$4\,arcsec$^2$ around this centroid-position for a compact PRS, but the highest peak we find is 0.072\,mJy (<5$\sigma$), still more than 3 times fainter than APTF~J230904+484222. This continuum radio source thus likely reflects star formation in the host galaxy, similar to the case of \rsixtyseven \citep{nimmo_2022_apjl}. Using the 1.4\,GHz luminosity-to-SFR relation \citep{Murphy+11}, a star-formation rate of about 2.5\,M${_{\odot}}$\,yr$^{-1}$ is required to explain the observed radio flux density of APTF~J230904+484222. This radio luminosity inferred SFR is consistent with the SFR of $\gtrsim$\,0.1\,M${_{\odot}}$\,yr$^{-1}$ derived from H$\alpha$ observations and is $\sim$3.5 times less than the star-formation rate inferred from radio observations of the host of \rsixtyseven \citep{dong_2023_arxiv}. 

If the PRSs associated with \rone and \ronetwin are in fact hyper-nebulae, powered by the central active magnetar \citep{sridhar_2022_apj}, the lack of a PRS in the case of \frb is particularly surprising, given how active this source is.  The absence of a PRS is, however, consistent with the stable and near-zero RM that suggests a non-turbulent and clean local environment \citep[e.g.,][]{feng_2023_arxiv}.

We strongly encourage multi-wavelength observations of FRB sources that are outliers in terms of proximity or activity, such as \frb and the others FRBs that have been localised to milliarcsecond precision. These observations can be a powerful means of characterizing the local environments of FRBs, and detailed studies such as these complement studies that provide less detailed information for a larger number of sources. Currently, with \textit{HST} it is possible to compare positions at the $\sim10$\,mas level ($\sim10$\% of the point spread function width), so there remains much to be gained from precision localisations. Scheduled to commence observations within a decade, the ELT will provide 5\,mas resolution, enabling an even stronger optical synergy to milliarcsecond localisations in the radio regime.  

\section*{Acknowledgements}

We thank the directors and staff at the various participating antenna stations for allowing us to use their facilities and running the observations. The European VLBI Network (EVN) is a joint facility of independent European, African, Asian, and North American radio astronomy institutes. Scientific results from data presented in this publication are derived from EVN project code EK051. Research by the AstroFlash group at the University of Amsterdam, ASTRON, and JIVE is supported in part by a Dutch Research Council (NWO) Vici grant (PI: Hessels; VI.C.192.045). SB is supported by an NWO Veni fellowship (VI.Veni.212.058). BM acknowledges financial support from the State Agency for Research of the Spanish Ministry of Science and Innovation under grant PID2019-105510GB-C31/AEI/10.13039/501100011033, PID2022-136828NB-C41 and through the Unit of Excellence Mar\'ia de Maeztu 2020--2023 award to the Institute of Cosmos Sciences (CEX2019-00918-M). FK acknowledges support from Onsala Space Observatory for the provisioning of its facilities/observational support. MB is a McWilliams fellow and an International Astronomical Union Gruber fellow. MB also receives support from the McWilliams seed grant. The Onsala Space Observatory national research infrastructure is funded through Swedish Research Council grant No. 2017-00648. e-MERLIN is a National Facility operated by the University of Manchester at Jodrell Bank Observatory on behalf of STFC, part of UK Research and Innovation. The Medicina and Noto radio telescopes are funded by the Italian Ministry of University and Research (MUR) and are operated as a National Facility by the National Institute for Astrophysics (INAF). Part of the research activities described in this paper were carried out with the contribution of the NextGenerationEU funds within the National Recovery and Resilience Plan (PNRR), Mission 4 - Education and Research, Component 2 - From Research to Business (M4C2), Investment Line 3.1 - Strengthening and creation of Research Infrastructures, Project IR0000026 – Next Generation Croce del Nord. This work is based in part on observations carried out using the 32-m radio telescope operated by the Institute of Astronomy of the Nicolaus Copernicus University in Toru\'n (Poland) and supported by a Polish Ministry of Science and Higher Education SpUB grant. This project has received funding from the European Union's Horizon 2020 research and innovation programme under grant agreements 730562 (RadioNet) and 101004719 (OPTICON-RadioNet Pilot).

\section*{Data Availability}

The EVN observations are available on the EVN Data Archive at JIVE\footnote{\url{http://archive.jive.nl/}} under project codes EK051G and EK051H. The relevant code and data products for this work will be uploaded on Zenodo at the time of publication.



\bibliographystyle{mnras}
\bibliography{frb} 




\appendix

\section{Burst properties}

Properties of the bursts detected in our PRECISE observations are reported in Table~\ref{tab:burst_properties}.

\onecolumn
\scriptsize
\begin{longtable}{llccc}
\caption{Burst properties}\\
\hline
		Burst ID & MJD$^a$       & Temporal width    & Fluence       & Spectral Energy Density    \\
               &       & (ms)$^b$          &  (Jy\,ms)$^{c,d}$ & (10$^{29}$ erg\,Hz$^{-1}$)$^{d}$       \\
		\hline
  \endhead
            G001 & 59874.01351809 & 4.4 $\pm$ 0.5 & 29 & 4.0 \\
            G002 & 59874.01727595 & 7.4 $\pm$ 0.7 & 31 & 7.3 \\
            G003 & 59874.01747271 & 7.2 $\pm$ 0.3 & 68 & 13.2 \\
            G004 & 59874.01793008 & 12.2 $\pm$ 0.8 & 61 & 14.6 \\
            G005 & 59874.01811878 & 0.99 $\pm$ 0.03 & 45 & 9.3 \\
            G006 & 59874.02032330 & 2.8 $\pm$ 0.1 & 64 & 13.6 \\
            G007 & 59874.02489297 & 4.5 $\pm$ 0.5 & 25 & 4.7 \\
            G008 & 59874.02843250 & 9.8 $\pm$ 1.2 & 31 & 5.3 \\
            G009 & 59874.03745237 & 9.7 $\pm$ 0.7 & 51 & 11.8 \\
            G010 & 59874.03786912 & 8.6 $\pm$ 1.0 & 38 & 4.1 \\
            G011 & 59874.03875683 & 4.4 $\pm$ 0.8 & 13 & 2.9 \\
            G012 & 59874.04316232 & 4.1 $\pm$ 0.4 & 22 & 4.0 \\
            G013 & 59874.04664155 & 4.2 $\pm$ 0.5 & 20 & 4.9 \\
            G014 & 59874.05413382 & 6.4 $\pm$ 0.6 & 32 & 6.5 \\
            G015 & 59874.06219702 & 4.7 $\pm$ 0.1 & 89 & 17.7 \\
            G016 & 59874.06233498 & 5.6 $\pm$ 0.4 & 44 & 6.9 \\
            G017 & 59874.06245408 & 12.1 $\pm$ 0.7 & 68 & 15.1 \\
            G018 & 59874.06263542 & 6.5 $\pm$ 0.6 & 31 & 6.2 \\
            G019 & 59874.06968019 & 13.4 $\pm$ 1.2 & 47 & 9.6 \\
            G020 & 59874.07035018 & 7.6 $\pm$ 0.4 & 52 & 10.8 \\
            G021 & 59874.07224331 & 5.3 $\pm$ 0.5 & 25 & 5.6 \\
            G022 & 59874.07548403 & 8.2 $\pm$ 0.8 & 37 & 5.9 \\
            G023 & 59874.08625452 & 7.6 $\pm$ 1.2 & 17 & 3.8 \\
            G024 & 59874.08928933 & 7.8 $\pm$ 0.4 & 62 & 13.5 \\
            G025 & 59874.09349927 & 4.0 $\pm$ 0.4 & 23 & 4.9 \\
            G026 & 59874.09427777 & 10.0 $\pm$ 0.7 & 61 & 11.7 \\
            G027 & 59874.09637159 & 3.6 $\pm$ 0.3 & 31 & 5.6 \\
            G028 & 59874.09921041 & 7.3 $\pm$ 0.7 & 28 & 6.6 \\
            G029 & 59874.09978949 & 4.3 $\pm$ 0.2 & 46 & 9.5 \\
            G030 & 59874.10045871 & 9.8 $\pm$ 0.8 & 43 & 8.8 \\
            G031 & 59874.10747114 & 7.5 $\pm$ 0.9 & 23 & 5.6 \\
            G032 & 59874.11355427 & 7.2 $\pm$ 0.3 & 102 & 18.6 \\
            G033 & 59874.11922025 & 6.5 $\pm$ 0.5 & 46 & 8.0 \\
            G034 & 59874.11991756 & 3.9 $\pm$ 0.2 & 64 & 11.5 \\
            G035 & 59874.12232641 & 5.3 $\pm$ 0.5 & 30 & 5.0 \\
            G036 & 59874.12390893 & 8.0 $\pm$ 0.5 & 48 & 11.6 \\
            G037 & 59874.13620542 & 8.5 $\pm$ 0.6 & 50 & 9.8 \\
            G038 & 59874.14528686 & 9.4 $\pm$ 0.5 & 78 & 11.3 \\
            G039 & 59874.15568159 & 6.1 $\pm$ 0.7 & 33 & 5.0 \\
            G040 & 59874.16142370 & 9.9 $\pm$ 0.5 & 73 & 12.4 \\
            G041 & 59874.16178252 & 4.0 $\pm$ 0.2 & 67 & 8.9 \\
            G042 & 59874.16582486 & 2.2 $\pm$ 0.2 & 21 & 2.5 \\
            G043 & 59874.18358549 & 11.1 $\pm$ 0.5 & 97 & 16.4 \\
            G044 & 59874.18760978 & 8.5 $\pm$ 0.4 & 78 & 14.0 \\
            G045 & 59874.19445752 & 5.4 $\pm$ 0.4 & 38 & 6.8 \\   
		\hline
            H001 & 59876.88324165 & 6.0 $\pm$ 0.4 & 40 & 7.2 \\
            H002 & 59876.88367760 & 2.6 $\pm$ 0.1 & 47 & 10.4 \\
            H003 & 59876.89137390 & 5.3 $\pm$ 0.4 & 44 & 7.3 \\
            H004 & 59876.89478408 & 4.0 $\pm$ 0.1 & 76 & 14.6 \\
            H005 & 59876.89971573 & 11.9 $\pm$ 0.5 & 102 & 21.1 \\
            H006 & 59876.89978801 & 4.3 $\pm$ 0.1 & 79 & 15.6 \\
            H007 & 59876.90039785 & 11.1 $\pm$ 0.4 & 117 & 25.4 \\
            H008 & 59876.90169720 & 4.5 $\pm$ 0.1 & 101 & 21.5 \\
            H009 & 59876.90958331 & 5.7 $\pm$ 0.2 & 94 & 22.6 \\
            H010 & 59876.91101023 & 5.7 $\pm$ 0.3 & 58 & 9.8 \\
            H011 & 59876.91463753 & 10.3 $\pm$ 0.4 & 88 & 21.1 \\
            H012 & 59876.91553057 & 12.3 $\pm$ 0.3 & 136 & 32.5 \\
            H013 & 59876.91577982 & 6.0 $\pm$ 0.2 & 79 & 16.3 \\
            H014 & 59876.91658368 & 7.6 $\pm$ 0.1 & 167 & 37.1 \\
            H015 & 59876.92139371 & 0.62 $\pm$ 0.04 & 16 & 3.0 \\
            H016 & 59876.92613409 & 3.7 $\pm$ 0.3 & 29 & 4.5 \\
            H017 & 59876.92625989 & 5.9 $\pm$ 0.5 & 35 & 7.7 \\
            H018 & 59876.92762306 & 4.2 $\pm$ 0.3 & 37 & 5.9 \\
            H019 & 59876.93432863 & 2.0 $\pm$ 0.3 & 12 & 2.0 \\
            H020 & 59876.94064355 & 5.0 $\pm$ 0.1 & 111 & 21.3 \\
            H021 & 59876.94137123 & 9.6 $\pm$ 0.3 & 147 & 25.8 \\
            H022 & 59876.94460382 & 7.9 $\pm$ 0.5 & 60 & 8.9 \\
            H023 & 59876.94530712 & 4.3 $\pm$ 0.3 & 44 & 8.4 \\
            H024 & 59876.94593990 & 10.6 $\pm$ 0.3 & 164 & 37.8 \\
            H025 & 59876.94619106 & 4.2 $\pm$ 0.6 & 20 & 3.1 \\
            H026 & 59876.94633808 & 7.1 $\pm$ 0.4 & 68 & 12.7 \\
            H027 & 59876.94734039 & 10.4 $\pm$ 0.4 & 111 & 17.3 \\
            H028 & 59876.95002759 & 4.7 $\pm$ 0.5 & 27 & 5.1 \\
            H029 & 59876.95047486 & 5.0 $\pm$ 0.6 & 20 & 3.9 \\
            H030 & 59876.95165196 & 3.7 $\pm$ 0.1 & 330 & 75.7 \\
            H031 & 59876.95273572 & 6.2 $\pm$ 0.4 & 43 & 7.7 \\
            H032 & 59876.96054167 & 4.2 $\pm$ 0.3 & 33 & 7.4 \\
            H033 & 59876.96082704 & 7.9 $\pm$ 0.7 & 40 & 7.8 \\
            H034 & 59876.96142198 & 7.6 $\pm$ 0.5 & 46 & 9.4 \\
            H035 & 59876.96154518 & 9.1 $\pm$ 0.4 & 77 & 15.3 \\
            H036 & 59876.96356728 & 6.1 $\pm$ 0.3 & 64 & 14.4 \\
            H037 & 59876.96613912 & 6.8 $\pm$ 0.3 & 79 & 16.7 \\
            H038 & 59876.96640209 & 3.9 $\pm$ 0.1 & 89 & 18.7 \\
            H039 & 59876.96936874 & 5.1 $\pm$ 0.3 & 52 & 11.1 \\
            H040 & 59876.97106987 & 8.5 $\pm$ 1.2 & 29 & 3.6 \\
            H041 & 59876.97160880 & 7.7 $\pm$ 0.4 & 62 & 13.5 \\
            H042 & 59876.97435577 & 9.0 $\pm$ 0.5 & 56 & 12.4 \\
            H043 & 59876.97603178 & 6.6 $\pm$ 0.3 & 59 & 13.5 \\
            H044 & 59876.97648147 & 5.6 $\pm$ 0.2 & 72 & 12.1 \\
            H045 & 59876.98304796 & 4.8 $\pm$ 0.3 & 34 & 7.4 \\
            H046 & 59876.98323034 & 4.5 $\pm$ 0.2 & 52 & 10.6 \\
            H047 & 59876.98348756 & 4.3 $\pm$ 0.2 & 71 & 16.3 \\
            H048 & 59876.98513941 & 34.0 $\pm$ 0.2 & 54 & 11.1 \\
            H049 & 59876.98526088 & 9.6 $\pm$ 1.1 & 31 & 6.3 \\
            H050 & 59876.98593060 & 5.1 $\pm$ 0.4 & 28 & 6.8 \\
            H051 & 59876.98819198 & 4.0 $\pm$ 0.4 & 25 & 4.7 \\
            H052 & 59876.99315217 & 8.3 $\pm$ 0.5 & 61 & 11.0 \\
            H053 & 59876.99340664 & 9.5 $\pm$ 1.2 & 27 & 5.5 \\
            H054 & 59876.99357744 & 6.9 $\pm$ 0.3 & 57 & 13.7 \\
            H055 & 59876.99559429 & 8.8 $\pm$ 0.6 & 49 & 9.7 \\
            H056 & 59876.99899468 & 4.4 $\pm$ 0.5 & 25 & 4.2 \\
            H057 & 59876.99935655 & 10.6 $\pm$ 0.8 & 54 & 12.4 \\
            H058 & 59876.99984536 & 7.7 $\pm$ 0.5 & 70 & 7.3 \\
            H059 & 59877.00043210 & 7.8 $\pm$ 0.8 & 37 & 6.1 \\
            H060 & 59877.00097951 & 8.8 $\pm$ 0.6 & 52 & 8.0 \\
            H061 & 59877.00835109 & 5.6 $\pm$ 0.4 & 43 & 7.1 \\
            H062 & 59877.00893303 & 4.5 $\pm$ 0.3 & 34 & 7.4 \\
            H063 & 59877.00906912 & 8.2 $\pm$ 1.3 & 21 & 3.6 \\
            H064 & 59877.00966560 & 4.6 $\pm$ 0.2 & 67 & 13.9 \\
            H065 & 59877.01003499 & 7.3 $\pm$ 0.4 & 53 & 11.7 \\
            H066 & 59877.01018686 & 5.1 $\pm$ 0.3 & 45 & 7.9 \\
            H067 & 59877.01205399 & 6.6 $\pm$ 0.9 & 26 & 3.7 \\
            H068 & 59877.01214328 & 4.7 $\pm$ 0.5 & 23 & 4.1 \\
            H069 & 59877.01324536 & 6.9 $\pm$ 0.5 & 40 & 7.9 \\
            H070 & 59877.01413124 & 7.0 $\pm$ 0.2 & 98 & 21.1 \\
            H071 & 59877.01801725 & 8.2 $\pm$ 0.5 & 50 & 8.7 \\
            H072 & 59877.02245511 & 3.3 $\pm$ 0.5 & 22 & 1.8 \\
            H073 & 59877.02286325 & 1.4 $\pm$ 0.1 & 19 & 4.6 \\
            H074 & 59877.03150464 & 9.7 $\pm$ 0.4 & 94 & 20.5 \\
            H075 & 59877.03195097 & 7.4 $\pm$ 0.8 & 44 & 5.1 \\
            H076 & 59877.03314940 & 4.4 $\pm$ 0.3 & 41 & 8.2 \\
            H077 & 59877.03426822 & 6.9 $\pm$ 0.8 & 29 & 5.7 \\
            H078 & 59877.03454198 & 8.0 $\pm$ 0.5 & 51 & 11.4 \\
            H079 & 59877.03641293 & 4.4 $\pm$ 0.6 & 20 & 2.8 \\
            H080 & 59877.03723562 & 7.6 $\pm$ 0.7 & 43 & 5.2 \\
            H081 & 59877.03799905 & 8.3 $\pm$ 0.3 & 79 & 19.0 \\
            H082 & 59877.03815510 & 6.5 $\pm$ 0.6 & 41 & 6.0 \\
            H083 & 59877.04259124 & 15.2 $\pm$ 0.2 & 429 & 102.8 \\
            H084 & 59877.0443353 & 4.4 $\pm$ 0.6 & 23 & 3.1 \\
            H085 & 59877.04472092 & 7.3 $\pm$ 0.5 & 53 & 9.2 \\
            H086 & 59877.04932255 & 6.6 $\pm$ 0.4 & 52 & 8.9 \\
            H087 & 59877.04953494 & 7.3 $\pm$ 0.5 & 44 & 8.0 \\
            H088 & 59877.05595215 & 7.1 $\pm$ 0.9 & 29 & 4.2 \\
            H089 & 59877.05718630 & 15.7 $\pm$ 1.4 & 54 & 9.5 \\
            H090 & 59877.05994232 & 3.6 $\pm$ 0.5 & 19 & 2.9 \\
            H091 & 59877.06011452 & 7.8 $\pm$ 0.4 & 68 & 15.4 \\
            H092 & 59877.06108046 & 12.4 $\pm$ 0.4 & 150 & 33.2 \\
            H093 & 59877.06118988 & 3.5 $\pm$ 0.5 & 17 & 3.3 \\
            H094 & 59877.06124699 & 4.0 $\pm$ 0.2 & 58 & 11.3 \\
            H095 & 59877.06140918 & 11.9 $\pm$ 1.1 & 43 & 8.2 \\
            H096 & 59877.06188390 & 6.5 $\pm$ 0.7 & 32 & 5.1 \\
            H097 & 59877.06229430 & 4.9 $\pm$ 0.4 & 38 & 7.4 \\
            H098 & 59877.06460628 & 8.4 $\pm$ 0.5 & 53 & 11.6 \\
            H099 & 59877.06650560 & 2.7 $\pm$ 0.3 & 39 & 1.8 \\
            H100 & 59877.06930657  & 9.0 $\pm$ 0.3 & 99 & 20.7 \\
            H101 & 59877.07478215  & 7.2 $\pm$ 0.6 & 64 & 5.2 \\
            H102 & 59877.07574649  & 4.8 $\pm$ 0.2 & 77 & 17.6 \\
            H103 & 59877.07668063  & 4.9 $\pm$ 0.5 & 28 & 4.2 \\
            H104 & 59877.07901416  & 7.4 $\pm$ 0.5 & 48 & 11.0 \\
            H105 & 59877.07957926  & 6.1 $\pm$ 0.7 & 28 & 4.0 \\

        \hline
\\

  \label{tab:burst_properties}      
\end{longtable}	
\noindent$^a$ The time-of-arrival of the burst at the solar system barycenter in TDB, corrected to infinite frequency for a DM of 219.37\,pc\,cm$^{-3}$ and using a DM constant of 1/(2.41$\times10^{-4}$) MHz$^2$\,pc$^{-1}$\,cm$^3$\,s.  \\ 
$^b$ FWHM of 1D Gaussian fit. \\
$^c$ Measured over the spectral extent of the burst. \\
$^d$ The estimated uncertainty is approximately 20 per cent due to uncertainty in the system equivalent flux density (SEFD).
  
\normalsize
\section{Individual burst localisation accuracy}

\subsection{Per Epoch Analysis}
\label{app:perepoch}

\begin{figure*}
\centering
    \begin{subfigure}[b]{0.8\textwidth}
        \centering
        \includegraphics[width=1\textwidth]{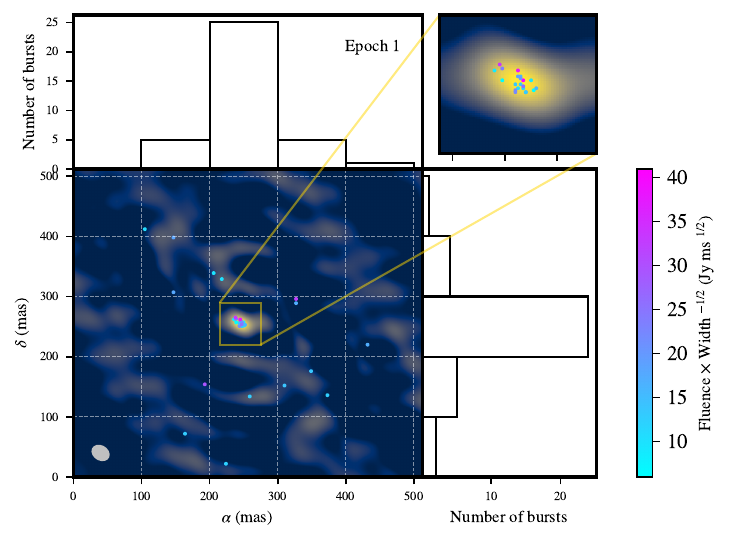}
    \end{subfigure}
    \begin{subfigure}[b]{0.8\textwidth}
        \centering
        \includegraphics[width=1\textwidth]{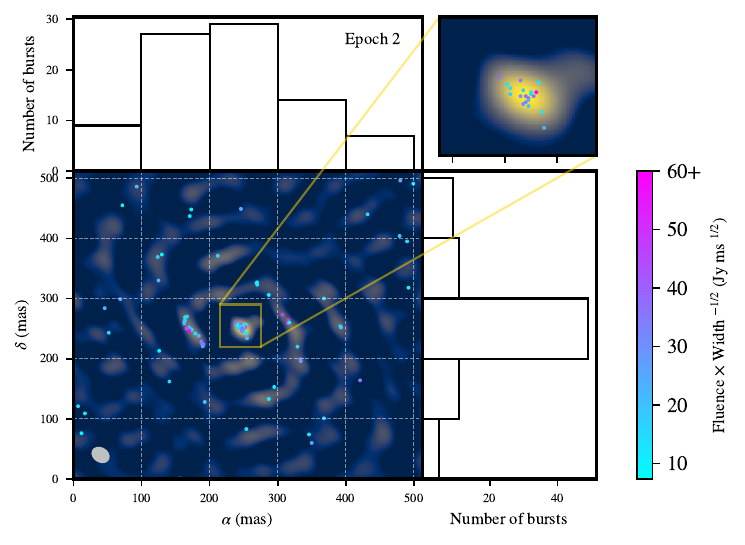}
    \end{subfigure}
    \caption{The position of the peak S/N of the dirty maps of individual bursts, compared to the best-known FRB position for Epoch~1 (top) and Epoch~2 (bottom). The points have been coloured according to a detection metric that is defined as fluence of the burst divided by the square root of the temporal width of the burst. The uncertainty of individual burst positions is underestimated, since only the peak value of the dirty map is used here for illustrative purposes. The true uncertainty of individual burst positions is more precisely known from taking into account the sidelobes. Bursts that occur during times when the calibration solutions are less robust are excluded from this analysis.}     
    \label{fig:individual_peaks}
\end{figure*}

\cite{marcote_2017_apjl} and \cite{nimmo_2022_apjl} investigated the astrometry of single burst localisations. Here we undertake a similar analysis. Figure~\ref{fig:individual_peaks} shows the peak position of the dirty maps of individual bursts overplotted on the dirty map of the combined visibilities of the 45 bursts in Epoch~1 and 105 bursts in Epoch~2. The data points have been coloured according to a detection metric that is defined as fluence of the burst divided by the square root of the temporal width of the burst \citep{marcote_2017_apjl}.
For the sake of brevity in the remainder of the discussion, when referring to the peak position of the dirty map of an individual burst, we will simply call it the individual peak positions. We observe that the individual peak positions are scattered around the best position (found by combining the visibilities of all bursts for an epoch). The offset between the individual peak positions and the best position can be as much as a few hundred milliarcseconds. We note that using the peak position of the dirty maps of a single burst to determine the localisation region is not ideal. As discussed in \cite{nimmo_2022_apjl}, the peak levels of the side lobes in the dirty maps are on average $>\,97$ per cent and in particularly bad cases where the bursts are faint and $uv$-coverage is poor it can be even higher than the main sidelobe. While brighter bursts that have a higher detection metric tend to have smaller offsets, the relation is not a simple linear one. Since the spectral extent of bursts differ, and the number of baselines differs for different parts of our observing band (see Figure~\ref{fig:freq_coverage}), the $uv$-coverage is dependent on where the burst falls within our observing frequency range. A very narrow-band burst might thus be at frequency ranges where the number of antennas is as little as two, and consequently due to the poor $uv$-coverage have a less accurate localisation.

An alternative approach to determining the position of a single burst has been presented in \cite{nimmo_2022_apjl}. The authors fit the cross-pattern observed in most dirty maps of individual bursts with two 2D Gaussians, and then fit the intersection of these Gaussians with another 2D Gaussian. This method has the advantage of smoothing over the prominent side lobes in these dirty maps, and thus provides a more robust localisation region than merely picking the highest peak in the dirty map. Fortunately, we have detected a very large number of bursts from \frb throughout our two observations, ensuring sufficient S/N to clearly identify the burst position and so we refrain from applying the aforementioned 2D-Gaussian fit method. Consequently, we do however note that the accuracy and precision of the single burst localisations shown here (peak of the dirty map) are significantly underestimated.

\bsp	
\label{lastpage}
\end{document}